\documentclass[11pt]{article}
\usepackage{instruction,epsfig}

\def\Journal#1#2#3#4{{#1} {\bf #2}, #3 (#4)}

\def\NIM{\em Nucl. Instrum. Methods}

\def\NPB{{\em Nucl. Phys.} B}
\def\PLB{{\em Phys. Lett.}  B}
\def\PRL{\em Phys. Rev. Lett.}
\def\PRD{{\em Phys. Rev.} D}

\def \JHEP{{\em JHEP}}
\def \JETP{{\em JETP Lett.}}

\def\mco{\multicolumn}

\def\ra{\rightarrow}

\def\ko{K^0}

\def\be{\begin{equation}}
\def\ee{\end{equation}}
\def\bea{\begin{eqnarray}}
\def\eea{\end{eqnarray}}

\begin{document}
\vspace*{4cm}
\title{Measurement of the $\pi\pi$ scattering length from a 
new structure in the $K^{\pm} \rightarrow 3\pi$ Dalitz plot}

\author{ Sergio Giudici \\ {\em on behalf of the NA48-CERN collaboration}}

\address{
Scuola Normale Superiore, 7 P.za dei Cavalieri \\
Pisa 56100 , Italy}

\maketitle\abstracts{We report here the results of a study of a partial 
sample of $2.773 \times 10^7$ $K^{\pm} \rightarrow \pi^\pm \pi^0 \pi^0$ decays
recorded in 2003, showing an anomaly in the $\pi^0\pi^0$ invariant mass 
($M_{00}$) distribution in the region around $M_{00}=2m_+$ where $m_+$ is the 
charged pion mass. This anomaly has never been seen in earlier experiments.
It can be interpreted as an effect of the charge exchanging scattering process 
$\pi^* \pi^- \rightarrow \pi^0 \pi^0$.} 

\section{Beams, detectors and reconstruction}
\begin{figure}
\psfig{figure=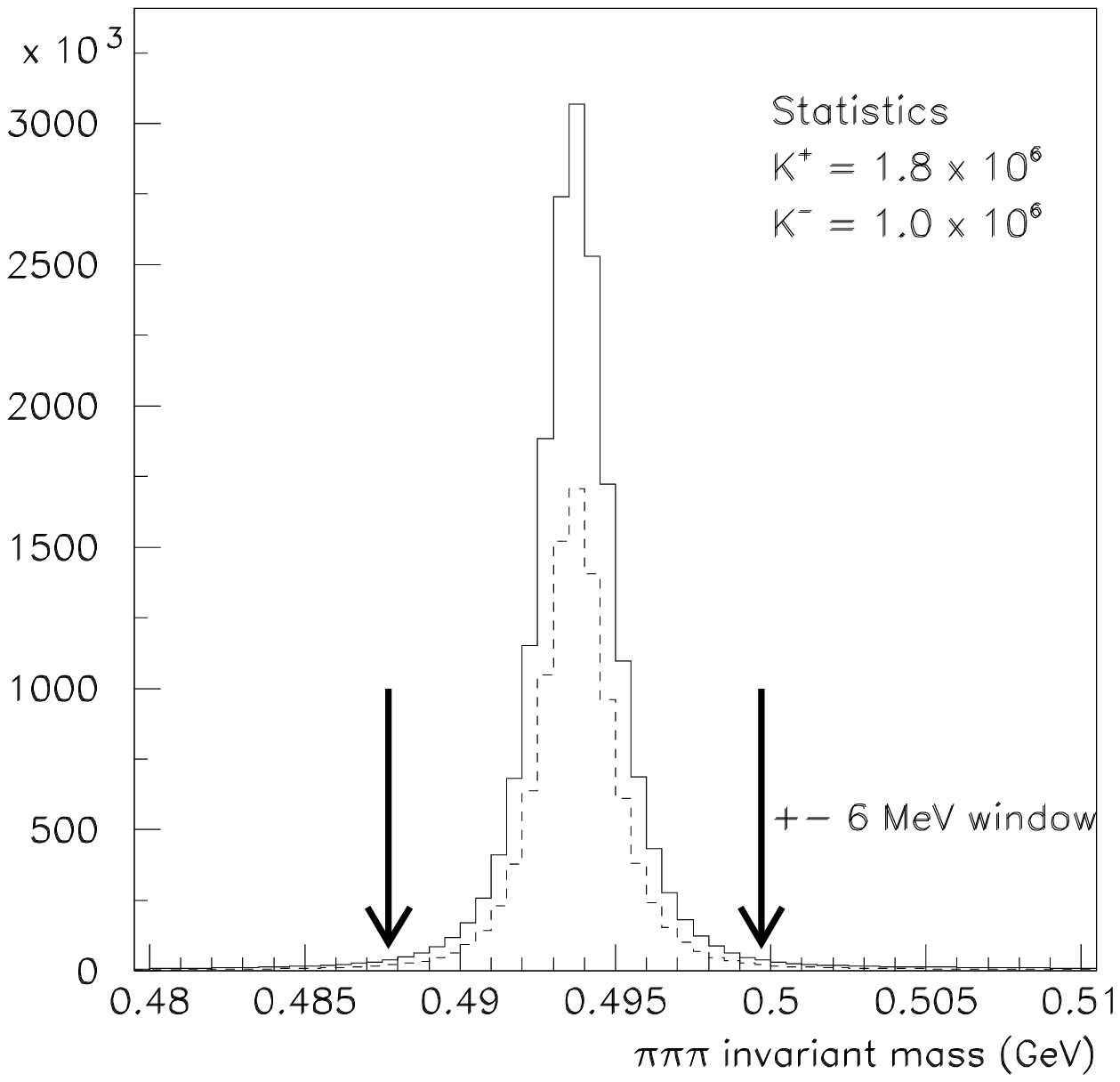,height=5.25cm}
\psfig{figure=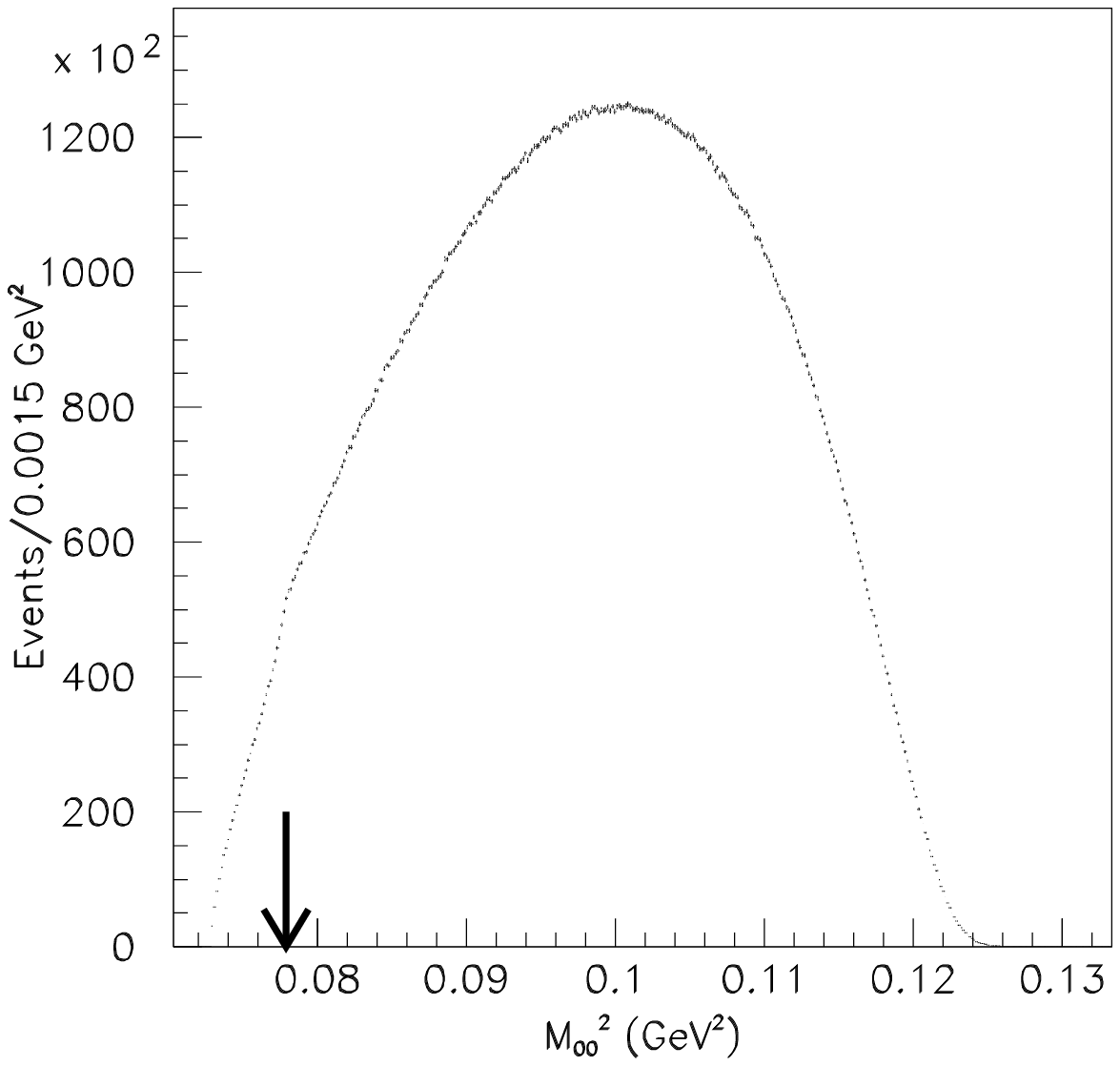,height=5.25cm}
\psfig{figure=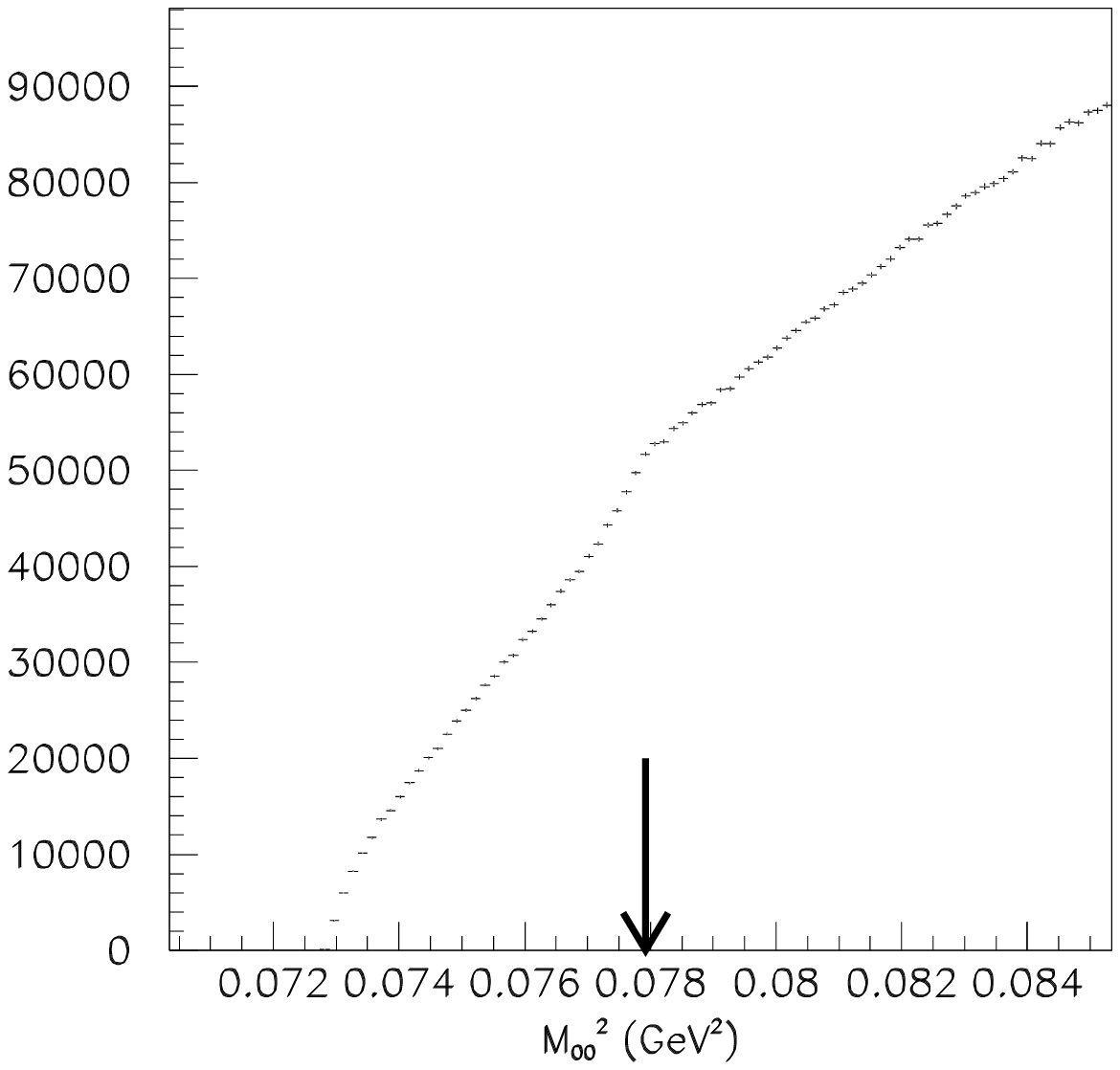,height=5.25cm}
\caption{(Left) Invariant mass distribution of $\pi^\pm \pi^0 \pi^0$ 
candidate events. (Middle) $\pi^0 \pi^0$ invariant mass squared distribution.
(Right) Zoom of the previous in the region
around the value $M_{00}= 2m_+$ indicated by the arrow.   
\label{fig:kmass}}
\end{figure}

During the years 2003-04, the experiment NA48 at CERN SPS has collected
a sample of fully reconstructed $\sim 2\times 10^8$ 
$K^{\pm} \rightarrow \pi^\pm \pi^0 \pi^0$ decays in order to study 
direct CP violation by comparing the Dalitz plot distributions of 
$K^+$ and $K^-$. 
The experiment makes use of two overlapping simultaneous focused kaon beams 
of opposite charge and momentum of $60~ GeV/c$ $\pm 3.8\%$, selected by 
a system of ``achromat'' magnets and collimators.
The two beams enter in a 114 m long decay volume and 
the $\pi^\pm \pi^0 \pi^0$ final state is reconstructed by combining the signals
coming from a spectrometer and from a liquid Krypton calorimeter (LKr) 
\cite{augustin}. 
The spectrometer, consisting of four drift chambers \cite {bede} and a dipole 
magnet located between the second and the third chamber, allows to track the charged pion
and to measure its momentum. The resolution is 
$\sigma(p)/p = 1.02\% \oplus 0.044\%p$, where $p$ is in GeV. 
The calorimeter is used to reconstruct the 
$\pi^0 \rightarrow \gamma \gamma$ decays. The resolution on $\gamma$ energy is
$\sigma(E)/E = 0.032/\sqrt{E} \oplus 0.09/E \oplus 0.0042$. 
The space resolution on the transverse coordinates can be parametrized as 
$\sigma_x = \sigma_y = 0.42/\sqrt{E} \oplus 0.06$ cm (E in GeV).
Events with at least one charged particle track and at least four energy 
clusters in the LKr are selected for further analysis. The distance between any
two $\gamma$-rays in the LKr is required to be larger than 10 cm and, 
in addition, the distance between each $\gamma$-ray and the impact point of
any track on LKr must exceed 15cm. Other cuts ensure full containment of the
electromagnetic shower in the LKr. The constraint of a common decay vertex for
the two $\pi^0$ is used to pair the four photons. From now on the correct
pairing is assumed to be (1,2) and (3,4). Momentum conservation and small angle
approximation allow to compute the distance D of the decay vertex to the LKr 
and the two $\pi^0$ invariant mass $M_{00}$ as:
\begin{equation}
\begin{array}{rcl}
D & = & \frac{1}{2m_0} \left(
\sqrt{E_1E_2 r_{12}^2} + \sqrt{E_3E_4 r_{34}^2} 
\right) \\
M_{00} & = & (1/D)\times \sqrt{\sum_{i,j = 1}^{4} E_iE_j r_{ij}^2} 
\\ & &
\end{array}\label{eq:kin}
\end{equation}
where $E_i,E_j$ are the energies of the i-th and j-th $\gamma$-ray and
$r_{ij}$ is their transverse distance at the LKr plane, $m_0$ is the $\pi^0$
mass. Note that both $D$ and $M_{00}$ computation involve only quantities 
measured by the LKr and no charge track parameters are used. This fact implies 
that even in the case of $\pi^{\pm} \rightarrow \mu^{\pm}$ decays in flight,
$M_{00}$ is correctly computed. Figure (\ref{fig:kmass},left) shows the 
invariant mass 
distribution of the system $\pi^\pm \pi^0 \pi^0$. The distribution is dominated
by the $K^{\pm}$ peak, as expected. 
Tails originate from $\pi^{\pm} \rightarrow \mu^{\pm}$ decays in flight.
The selection requires that the $\pi^\pm \pi^0 \pi^0$ invariant mass differs
from  the nominal $K^{\pm}$ mass quoted in the PDG by no more than $\pm 6$ MeV. 
The fraction of events with wrong $\gamma$-ray pairing is $\sim 0.25\%$ as 
estimated by a Montecarlo simulation. 
\section{$\pi^0 \pi^0$ invariant mass distribution}
Figure (\ref{fig:kmass},middle) and (\ref{fig:kmass},right) show the two neutral 
pions invariant squared mass $M_{00}^2$. The r.m.s of the $M_{00}^2$ resolution 
curve increases with $M_{00}^2$, varying between $\sim 0.0002 ~GeV^2$
and $\sim 0.001~GeV^2$ at the end of the allowed $M_{00}^2$ range.
This excellent resolution is the result of the intrinsic energy and 
spatial resolution of LKr. A sudden change in the slope (Cusp) near 
$M_{00}= 2m_+$ can be clearly seen. 
At this point the resolution on $M_{00}$ is 
$\sim 0.0003~GeV^2$. Such an anomaly has not been observed in 
previous experiments.  
\section{Interpretation of the cusp}
\begin{figure}
\begin{center}
\psfig{figure=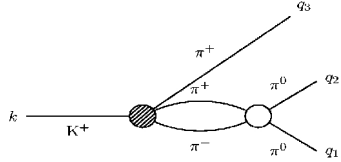,height=2.5cm}
\end{center}
\caption{1 loop diagram contributing to 
$K^{\pm} \rightarrow \pi^\pm \pi^0 \pi^0$.   
\label{fig:fey}}
\end{figure}

The observed sudden change of the slope suggests the presence of a threshold 
``Cusp'' effect from the decay $K^{\pm} \rightarrow \pi^\pm \pi^+ \pi^-$
 contributing to $K^{\pm} \rightarrow \pi^\pm \pi^0 \pi^0$ amplitude through
the charge exchange reaction $\pi^+ \pi^- \rightarrow \pi^0 \pi^0$.
This phenomenon has been recently discussed by Cabibbo \cite{cab1} who computed
the  $K^{\pm} \rightarrow \pi^\pm \pi^0 \pi^0$ amplitude
taking into account the 1-loop diagram shown in figure \ref{fig:fey}.
The diagram produces a discontinuity when the looping charged pions pair flips 
from off to on mass shell at the value $M_{00} = 2m_+$. 
The PDG parametrization for the  $K^{\pm} \rightarrow \pi^\pm \pi^0 \pi^0$
decay amplitude ${\cal M}_0$ is given by
\begin{equation}
{\cal M}_0 = 
\sqrt{(1 + \frac{1}{2}gu)^2+h^\prime u^2 }
\label{eq:tree}
\end{equation}
$u=(s_3-s_0)/m_+^2$ is the Lorentz-invariant variable where  
$s_i = (P_K -P_i)^2~(i=1,2,3)$, $P_K$ and $ P_i$ are 
the 4-momentum vectors respectively of the initial Kaon and of the three 
outgoing pions; $s_0 = (s_1+s_2+s_3)/3$ and of course $M_{00}^2 = s_3$.
In the Cabibbo theory, the diagram in figure \ref{fig:fey} is responsible 
for a new term in the amplitude 
${\cal M}_1 \propto (a_0-a_2)J$
proportional to the difference between 
the $I=0$ and $I=2$ S-wave $\pi\pi$ scattering lengths.
The term $J$ changes from real to imaginary at $M_{00} = (2m_+)$.
The distructive interference of ${\cal M}_0$ and ${\cal M}_1$ 
in the total amplitude ${\cal M} = {\cal M}_0 + {\cal M}_1$ is responsible 
for the cusp and for the apparent lack of event below the threshold.
More recently Cabibbo and Isidori \cite {cab2} have extended the amplitude   
calculation at 2-loops level suggesting a precise way to extract the $(a_0-a_2)$ parameter
from the data. 
The most recent theoretical prediction  
$(a_0-a_2)m_+ = 0.265 \pm 0.004$ is given in reference \cite{colangelo}.   
\section{Fit and results}
\begin{table}[here]
\begin{center}
\begin{tabular}{c c}
Model & $\chi^2 /ndf$ \\
\hline
\`{a} la PDG (full region) & 15000/139\\
\`{a} la PDG (above cusp)  & 240/198\\
Cabibbo 1-loop             & 463/149\\
Cabibbo-Isidori 2-loops    & 159/147\\
2-loops and pionium        & 154/146\\
\hline
\end{tabular}
\end{center}
\caption{fit against various models}
\label{tab:chi2}
\end{table}
Table \ref{tab:chi2} shows the $\chi^2$ obtained when the experimental 
$M_{00}^2$ spectrum is fitted against various model. The naive PDG description 
cannot account for the cusp structure observed and it gives a ``crazy'' 
$\chi^2$ although it seems to fit reasonably above the cusp. 
The Cabibbo 1-loop theory is still not adequate to fit the spectrum
while the 2-loop calculation is quite satisfactory. 

\begin{figure}
\begin{center}
\psfig{figure=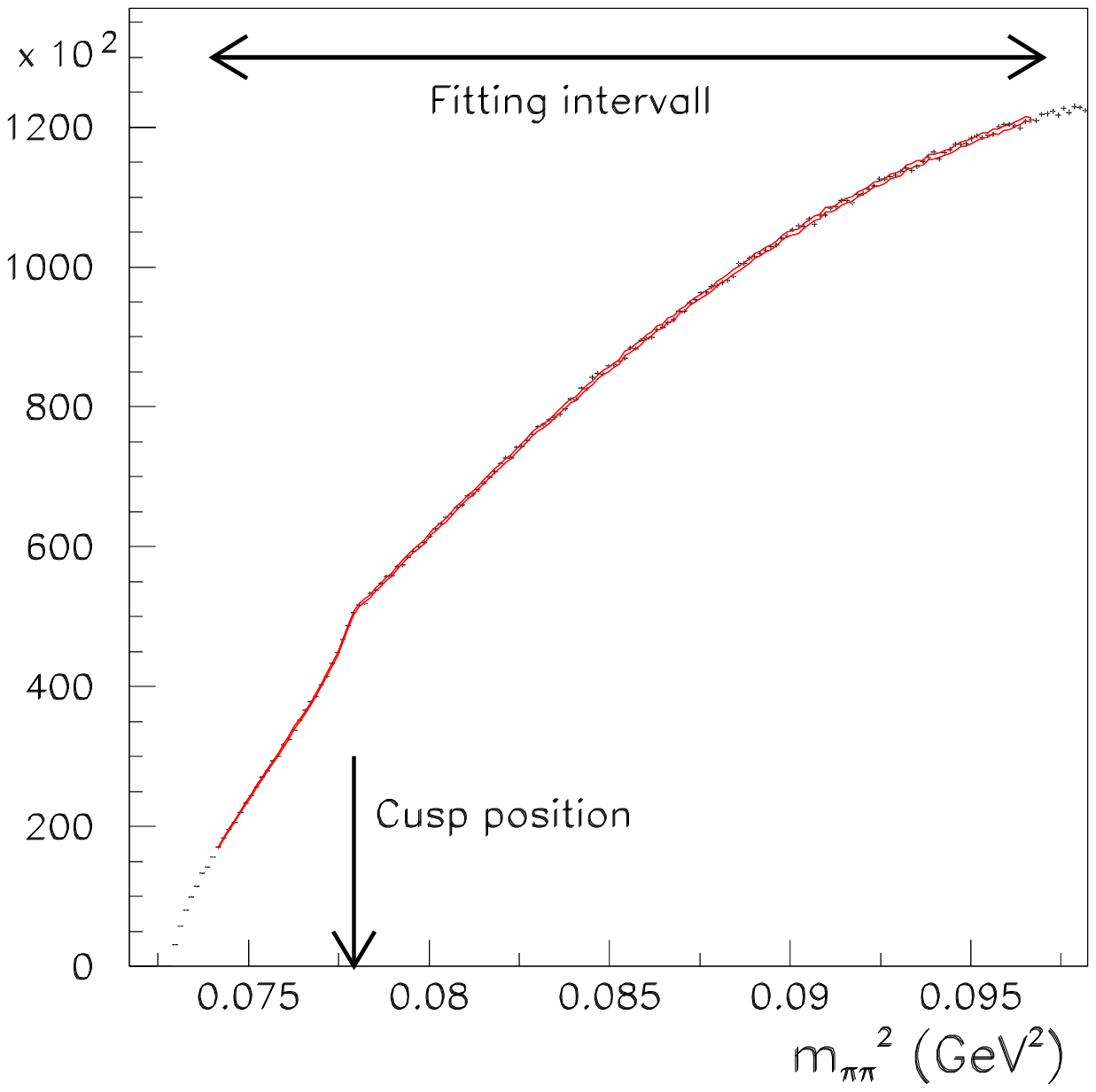,height=5.25cm}
\psfig{figure=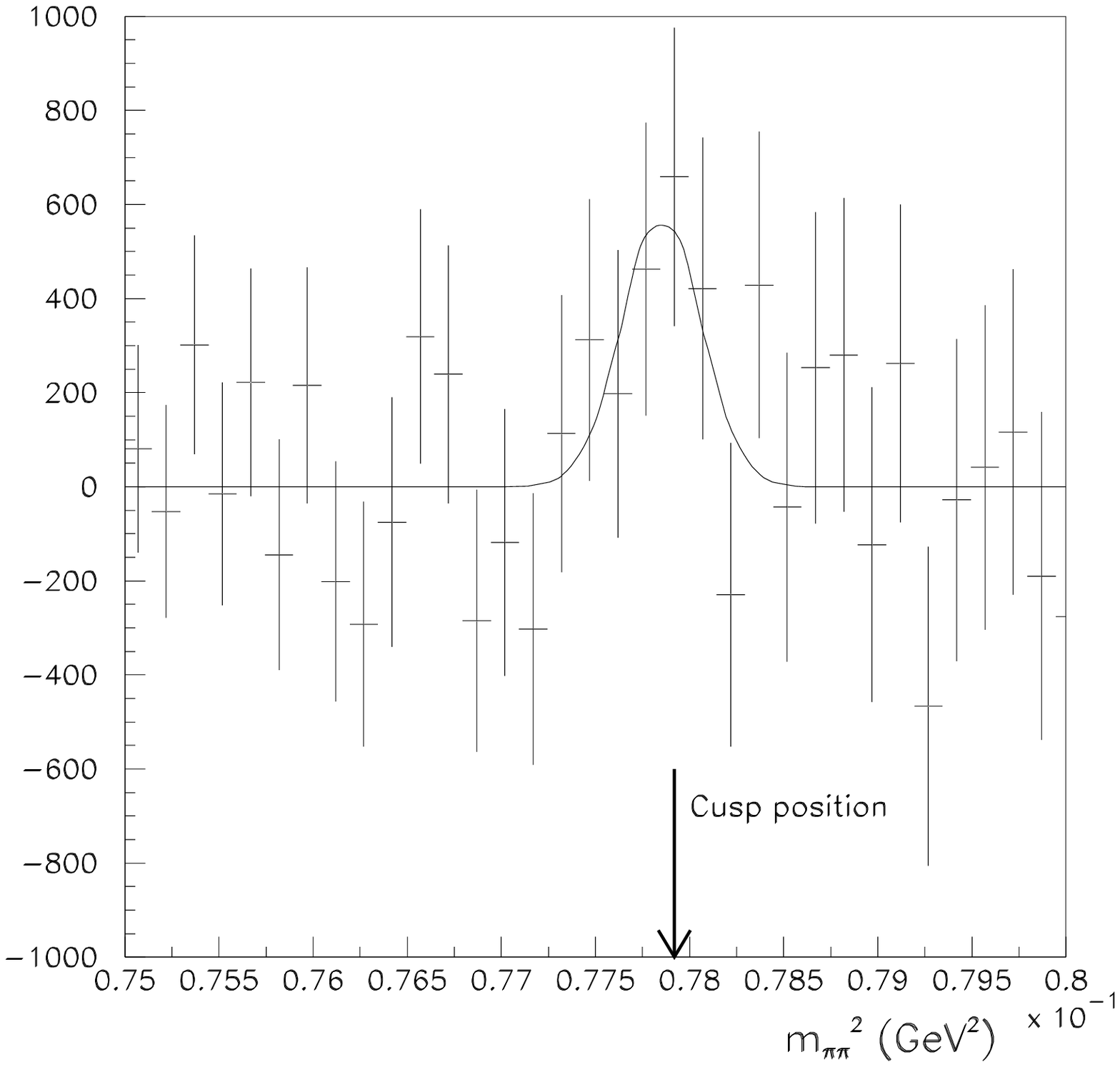,height=5.25cm}
\end{center}
\caption{(Left) Best fit to data. (Right) Excess of event at $M_{00}=2m_+$ 
interpreted as pionium formation.
\label{fig:pionium}}
\end{figure}
Figure \ref{fig:pionium} at left shows the best fit to data in the interval
$0.074<M_{00}<0.097 ~GeV^2$  obtained by slightly modifying the
Cabibbo-Isidori model to account for a small excess of event located
exactly at $M_{00}=2m_+$ on top of the observed cusp.
This excess can be interpreted as formation of pionium.  
The pionium is the electromagnetic bound state $\pi^+ \pi^-$ which is expected
to decay mainly in $\pi^0 \pi^0$ with a time constant $10^{-16}$ sec.
Figure \ref{fig:pionium} at right shows the distribution of the difference
between data and fit when the pionium contribution is set to zero in the fit.
A fixed amount of pionium derived by the estimation published in 
\cite{silagadze} has been used for the final fit.
The value found for the scattering length parameter is 
$(a_0-a_2)m_+ = 0.281 \pm 0.007$ where the error is only statistical.
 \begin{table}[here]
\begin{center}
\begin{tabular}{c c}
Source & Error \\
\hline
 skipping pionium & $\pm 0.008$\\
Cut on photon to track distance & $\pm0.004$\\
Dependence on vertex position & $\pm0.009$\\
$K^+/K^-$ difference & $\pm0.006$\\
\hline
total & $\pm 0.014$
\end{tabular}
\end{center}
\caption{Systematics error on $(a_0-a_2)m_+$}
\label{tab:syst}
\end{table}
A very conservative estimation of the systematic uncertainties is given 
in table \ref{tab:syst}. We tried to exclude 10 points around the cusp 
from the fit to reduce the sensitivity to the pionium component. Other checks
have been performed by changing the cut on the minimum distance 
between photons and charged pion at the LKR or by selecting events 
in different region of the decay vertex.   
As a preliminary result we quote the value
$$ (a_0 -a_2)m_+ = 0.281 \pm 0.007 (stat.) \pm 0.014 (syst.) \pm 0.014 (theo.)$$
The theoretical error, equal to the $5\%$ of the central value, has been suggested
by Cabibbo and Isidori as the achieved accuracy in their model.
\section {Conclusions} The $\pi^0 \pi^0$ invariant mass ($M_{00}$) measured from 
a sample of $2.773 \times 10^7$ $K^{\pm} \rightarrow \pi^\pm \pi^0 \pi^0$ fully 
reconstructed decays collected by NA48/2 experiment at the CERN-SPS shows an 
anomaly at $M_{00} = 2m_+$. This anomaly can be explained by a simple model 
based on contribution from the decay $K^{\pm} \rightarrow \pi^\pm \pi^+ \pi^-$
through the charge exchange reaction $\pi^+ \pi^- \rightarrow \pi^0 \pi^0$.
A high-statistics measurement of the $M_{00}$ distribution could be used therefore
to provide a precise determination of the $\pi \pi$ scattering parameter 
$(a_0-a_2)$. The expected increase of the event sample statistics 
by more than a factor 5 coming from the data collected in 2004 should provide 
a substantial reduction of the conservative systematics quoted here.
The quality of the data calls for an improved theory including  
isospin breaking effects and radiative corrections.

\end{document}